# Uniform e-beam irradiation-induced athermal straightening of axially curved amorphous SiO$_x$ nanowire

Jiangbin Su[1,2], Xianfang Zhu[1*]

1. China-Australia Joint Laboratory for Functional Nanomaterials & Physics Department, Xiamen University, Xiamen 361005, PR China

2. Experiment Center of Electronic Science and Technology, School of Mathematics and Physics, Changzhou University, Changzhou 213164, PR China

* Corresponding author. E-mail: zhux@xmu.edu.cn

**Abstract:** The reshaping of amorphous SiO$_x$ nanowires (a-SiO$_x$ NWs) as purely induced by uniform electron beam (e-beam) irradiation was *in-situ* studied at room temperature in transmission electron microscope. It was observed that the axially straight NW kept its perfect straight cylinder-like wire shape and demonstrated a sequential uniform radial shrinkage with the increase of irradiation time. In contrast, the axially curved NW turned straight quickly accompanied with a uniform axial shrinkage and a uniform radial expansion intriguingly. It is expected that such a study especially on the straightening of axially curved NW has important implications for the nanoscale processing and stability of future NW-based structures or devices. More importantly, the findings demonstrate that the traditional knock-on mechanism and electron beam heating effect are inadequate to explain these processes while our proposed nanocurvature effect and energetic beam-induced athermal activation effect obviously dominate the processes.

# 1. Introduction

As one of the important quasi-one dimensional blue photoluminescence nanomaterials, amorphous $SiO_x$ nanowires (a-$SiO_x$ NWs) have attracted much attention in the fields such as scanning near-field optical microscopes and integrated optical devices [1]. However, due to the difficulties in controllable synthesis or assembly, as-prepared a-$SiO_x$ NWs may not have ideal wire shapes such as straightness, diameter and length. Furthermore, such glassy silicon oxides are characterized by covalent and ionic bonding, and hence brittle materials. They do not exhibit ductility at room temperature to allow plastic deformation after their fabrication into structures or devices. It may thus limit the application of a-$SiO_x$ NWs in the related devices to a certain degree. Therefore, it is imperative and crucial to explore the precise and flexible reshaping of the NW structure at room temperature. Previous studies have demonstrated that energetic electron beam (e-beam) in transmission electron microscope (TEM) is a powerful tool which can not only *in-situ* observe the atomic NW structure but also athermally induce the NW processing [2-8]. Up till now, some e-beam induced or assisted nanoprocessings in axially straight a-$SiO_x$ NWs such as the elongation, shrinkage, cutting and necking have been reported [6-8], in which the length or diameter changes were realized. In spite of these, there is still no report found on the straightening of axially curved a-$SiO_x$ NWs as purely induced by e-beam irradiation in an electron microscope. Moreover, different from only positive nanocurvature uniformly distributed over the sidewall of the conventionally-studied axially straight NW, both positive and negative nanocurvature are non-uniformly distributed over the sidewall of the axially curved NW. It is expected that such axially curved NWs would cause much more intrinsic structural instability and thus greatly different e-beam-induced kinetic behavior. However, to our best knowledge, the nanocurvature effect of a-$SiO_x$ NW [9-10] and the athermal activation effect as induced by energetic beam [10-11], which are regarded as the key factors dominating the beam-induced structure changes and atom transportations, have not been studied in such axially curved NWs. In this regard, study on the straightening of axially curved a-SiO$_x$ NWs at room temperature also makes great sense to a further fundamental understanding of new nanophenomena and concepts in low dimensional nanostructures (LDNs).

## 2. Experimental

The a-SiO$_x$ NWs were grown by our improved chemical vapor deposition set-up where x is determined to be 2.3 [12]. They were well-dispersed in ethanol and then deposited onto the holey carbon film of microscopy grid for TEM studies. As-prepared TEM specimens were irradiated at room temperature and the structure changes and evolution of SiO$_x$ NWs were *in-situ* observed *via* a field-emission Tecnai F30 TEM operated at 300 kV. The irradiation was always targeted on axially curved or straight (for comparison) segment of single clean wire protruding into the open space of the holes in the carbon film of microscopy grid. Meanwhile, the two ends of the selected wires were lying on the carbon film surface and further fixed at ropes of a-SiO$_x$ NWs. In each irradiation, the current density at the specimen was kept at 1 A/cm$^2$ (electron flux: 6.25×10$^4$ nm$^{-2}$ s$^{-1}$) which was uniform over an area larger than the zone or NW observed. During the observation or taking a picture, the beam was spread to an around 100 times weaker intensity so that the corresponding irradiation effect can be minimized to a negligible degree and at the same time the image contrast can also be improved. Also note that during the electron irradiation, the beam was expected to heat the specimen by no more than a few degrees [5-6,13] due to its extremely large ratio of surface to volume and the dominant irradiation effect should be athermal. Therefore, it could be considered that the irradiated NW essentially remained at room temperature throughout the irradiation duration. In the experiments, the wire diameter was taken as an average value of several wire diameters which are measured at different representative positions across the wire from each micrograph; while the length and volume of the wire segment were measured and calculated between two black mark dots in (a) of Figs. 1-2 along the curved or straight wire axis. The two black dots were carefully marked at the locations of two feature points and by further checking the relative positions of their surrounding feature points on the wire surface (see the arrows in (a) of Figs. 1-2).

## 3. Results and Discussion

The sequential TEM micrographs in Fig. 1(a) show the typical *in-situ* structure changes and evolution of an axially straight a-SiO$_x$ NW segment during uniform e-beam irradiation with a current density at 1 A/cm$^2$. It was observed that the axially straight NW segment kept its perfect straight cylinder-like wire shape during the whole irradiation duration up to 2880 s. Furthermore, since the ends were both fixed, the NW segment could not shrink in the axis direction and thus only shrunk in its diameter to minimize the surface energy with reducing its surface area. As further quantified in Fig. 1(b), with the increase of irradiation time (or electron dose), the length almost kept a constant of 288 nm while the diameter decreased uniformly and slowly from 40.4 nm to 34.7 nm with an average radial shrinking rate of $2.0\times10^{-3}$ nm/s. Meanwhile, the volume of the NW segment decreased slowly from $3.68\times10^5$ nm$^3$ to $2.75\times10^5$ nm$^3$ with an average rate of $3.23\times10^1$ nm$^3$/s (see the inset in Fig. 1(b)). In contrast, as shown in Fig. 2(a), under the same irradiation conditions the axially curved NW segment of a circular arc-like shape turned straight intriguingly within the short duration of 130 s with its two ends nearly no moving or outward-protruding. Meanwhile, a notable uniform axial shrinking and radial thickening was occurring on the wire segment within the carbon film hole. As a result, the axially curved NW segment changed into an axially straight one quickly with a shorter length and a thicker diameter. As further quantitatively shown in Fig. 2(b), with the increase of irradiation time the length decreased quickly from 275.9 nm to 229.3 nm with an average axial shrinking rate of $3.6\times10^{-1}$ nm/s while the diameter increased much slowly from 26.9 nm to 28.0 nm with an average radial thickening rate of $8.5\times10^{-3}$ nm/s. Meanwhile, the volume of the NW segment decreased quickly from $1.57\times10^5$ nm$^3$ to $1.41\times10^5$ nm$^3$ with an average rate of $1.23\times10^2$ nm$^3$/s (see the inset in Fig. 2(b)), which is almost 4 times that in the axially straight NW case. It was further observed that even after the above evolution in the axially straight or curved NW proceeded for a while, it would stop immediately once the irradiation was suspended. This means that the process is predominately driven by an ultrafast irradiation-induced athermal activation rather than a slow beam heating-induced thermal activation. It also thus indicates a reshaping process of the axially curved NW as purely induced by the uniform e-beam irradiation. A similar irradiation on other axially curved and straight NW segments was repeated several times. We observed that the features of the structure changes and evolution were essentially the same as those shown in Figs. 1-2.

    As experimentally demonstrated above, the brittle a-SiO$_x$ NW segments exhibit remarkable athermal reshapings such as straightening, shrinking, or thickening at room temperature under the uniform e-beam irradiation. It was generally considered to be impossible to achieve such plastic deformations in glassy materials at room temperature. This is because the glass-transition temperature is as high as 1373 K and room temperature is too low for viscous flow to contribute significantly to the accommodation of the imposed deformation rate [7]. However, with the assistance of uniform e-beam irradiation, Zheng *et al* [7] have realized a superplastic elongation of axially straight silica NWs by tensile pulling; by a pure focused e-beam irradiation, we have achieved a locally-prolonged S-type deformation in an axially straight a-SiO$_x$ NW [6]; further by a pure uniform e-beam irradiation, as shown here, we have actualized a straightening, axial shrinking and radial thickening in axially curved

a-SiO$_x$ NWs. All of these demonstrate that the energetic e-beam can athermally soften the brittle amorphous silica or SiO$_x$ NW and cause remarkable reshapings with or even without the external assistance of tensile pulling.

Normally, people often resort to the traditional science concepts such as knock-on mechanism [13] and e-beam heating effect [2,14-15] to explain and predict the energetic beam purely induced nanophenomena. However, the knock-on mechanism and its related simulations have to be based on the equilibrium, symmetry, periodicity and linear nature of bulk crystalline structures, or its approximations. Therefore, the knock-on mechanism and the related simulations cannot differentiate the nanocurvature effect [9-10] of NWs, or account for the beam-induced athermal activation effect [10-11], both of which are of non equilibrium, amorphous-like, or nonlinear nature. This is especially true when we explain the effects associated with the fast straightening, shrinking and thickening of NW and surface atom adjustment or refilling at room temperature when size of the NW approaches the atomic distance at nanoscale and the beam energy deposition rate [8] is very fast. On the other hand, due to the extremely high ratio of surface to volume of NW at the nanoscale, the e-beam irradiation is expected to heat the specimen by no more than a few degrees [5-6,13], and the dominant irradiation effect should be athermal. Moreover, as we have mentioned above, the fact that the evolution will stop immediately once the irradiation is suspended also demonstrates an ultrafast irradiation-induced athermal activation effect. In fact, our previous research [6,8-11] has demonstrated that our proposed novel nanocurvature effect and energetic beam-induced athermal activation effect are universal phenomena and applicable in prediction or explanation of nanophenomena of LDNs. In the following, we attempt to reveal the above two effects on the reshapings of a-SiO$_x$ NW especially the axially curved a-SiO$_x$ NW under a uniform e-beam irradiation.

For the nanocurvature effect on an axially straight NW, we can suppose that, similar to the particle case [9-10], when the radius of the NW approaches its atomic bond length, a positive nanocurvature on the radially-curved wire surface will become appreciable. As schematically illustrated in Fig. 3(a), however, there is no curvature in the straight wire axis direction. In contrast, for an axially curved NW as shown in Fig. 3(b), besides the radial positive nanocurvature, there are a positive and a negative nanocurvature further produced respectively on the convex outer sidewall and the concave inner sidewall (similar to the cavity case [9-10]) along the curved wire axis. The positive nanocurvature would cause an additional tensile stress on the electron cloud structure of surface atoms on the convex wire sidewall. As a result, the vibration frequency of the surface atoms would be decreased and thus the "Debye temperature" (the concept which is borrowed to account for the phenomena in a LDN or an amorphous structure cannot be defined in the strict sense) would be lower and cause the convex wire sidewall to melt and the atoms therein to migrate or escape out. However, the negative nanocurvature would cause an additional compressive stress on the electron cloud structure of surface atoms on the concave wire sidewall. The compressive stress would lead to a speeding up of the vibration of the surface atoms and thus increase the "Debye temperature" and induce the concave wire sidewall to get other atoms. In this way, the positive or negative nanocurvature would cause intrinsic structural instabilities in the NW especially in the axially curved NW. This phenomenon was called nanocurvature effect or nanosize effect in a broad sense [9-10].

Although the nanocurvatures on a NW can cause structural instability, a further assistance from external excitation such as energetic beam irradiation is still needed to realize finally the ultrafast mass transportations and structure changes. This can be verified by the observation that the straightening, shrinking and thickening of NWs would stop immediately once the irradiation was suspended even after the structure changes started. In the present case of energetic e-beam irradiation in TEM, we can assume that, when the beam energy deposition rate of the incident energetic beam becomes very fast, there is no enough time for the deposited energy to transfer to thermal vibration energy of atoms within a single period of the vibration, and thus the mode of atom thermal vibration would be softened or the vibration of atoms would lose stability [10-11]. The as-induced soft mode or instability of atomic vibration can suppress the energy barrier or even make it totally disappear. In doing so, the irradiation would induced atom transportations such as the athermal diffusion (or even athermal plastic flow) or/and athermal evaporation (or even athermal ablation) of the wire atoms and thus the structure changes. This phenomenon was called beam-induced athermal activation (or beam-induced atomic vibration soft mode and instability) effect, or nanotime effect in a broad sense [10-11].

Moreover, as the curvature is uniform over the surface of the axially straight NW, under uniform e-beam irradiation the wire atoms can only evaporate uniformly from the wire surface (see Fig. 3(a)) to minimize the surface energy. As a result, it demonstrates a sequential uniform radial shrinkage with the increase of irradiation time. However, for the axially curved NW as shown in Fig. 3(b), the curvature distributed over the wire surface is not uniform along the curved wire axis. As driven by the positive and negative nanocurvature effects, the athermally activated or softened wire atoms on the convex sidewall of positive curvature would diffuse or flow to the opposite concave sidewall of negative curvature or preferentially evaporate into the vacuum (see Fig. 3(b)). In this way, the axially curved NW segment turns straight quickly along with a decrease in its length and an increase in its diameter. As shown in Fig. 2, the increase of wire diameter against the irradiation time seems less pronounced or to slow down. This can be attributed to the decreasing of curvature difference between the convex and the concave sidewall during the straightening process. In addition, relative to the axially straight NW, the faster decreasing of wire volume in the axially curved NW as mentioned above demonstrates a larger nanocurvature and a higher resulting atom evaporation rate.

## 4. Conclusions

In this work, we studied the *in-situ* reshapings of a-SiO$_x$ NWs as purely induced by uniform e-beam irradiation at room temperature in transmission electron microscope. It was observed that the axially straight NW kept its perfect straight cylinder-like wire shape and demonstrated a sequential uniform radial shrinkage with the increase of irradiation time. In contrast, the axially curved NW turned straight quickly accompanied with a uniform axial shrinkage and a uniform radial expansion intriguingly. The athermal diffusion and plastic flow of wire atoms in the axially curved NW indicates a direct experimental evidence for our predicted athermal activation effect as induced under uniform e-beam irradiation in condensed matter. Furthermore, the directional diffusion and flow of wire atoms from the convex wire sidewall to the nearby concave wire sidewall and the preferential evaporation of wire atoms on the convex wire sidewall demonstrate the effect of positive and negative nanocurvatures non-uniformly distributed in the axially curved NW. It is expected that such a study has important implications for the nanoscale processing and stability of future NW-based structures or devices. More importantly, the findings demonstrate that the current knock-on mechanism and electron beam heating effect are inadequate to explain these processes but our proposed nanocurvature effect and energetic beam-induced athermal activation effect obviously dominate the processes.

## Conflicts of interest

There are no conflicts of interest to declare.

## Acknowledgments

This work was supported by the NSFC project under grant no. 11574255, the Science and Technology

Plan (Cooperation) Key Project from Fujian Province Science and Technology Department under grant no. 2014I0016, and the National Key Basic Science Research Program (973 Project) under grant no. 2007CB936603.

**Figures:**

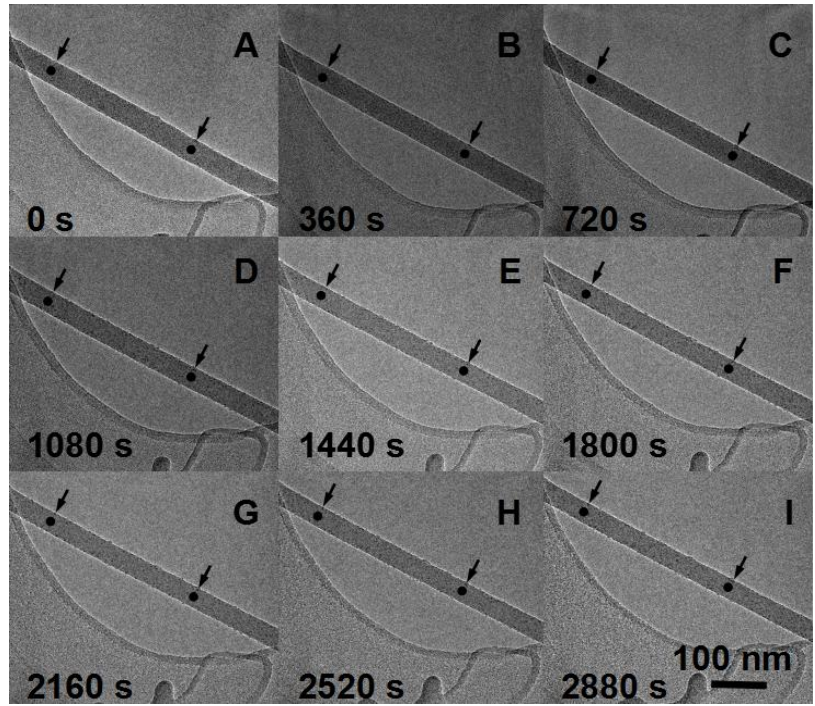

(a)

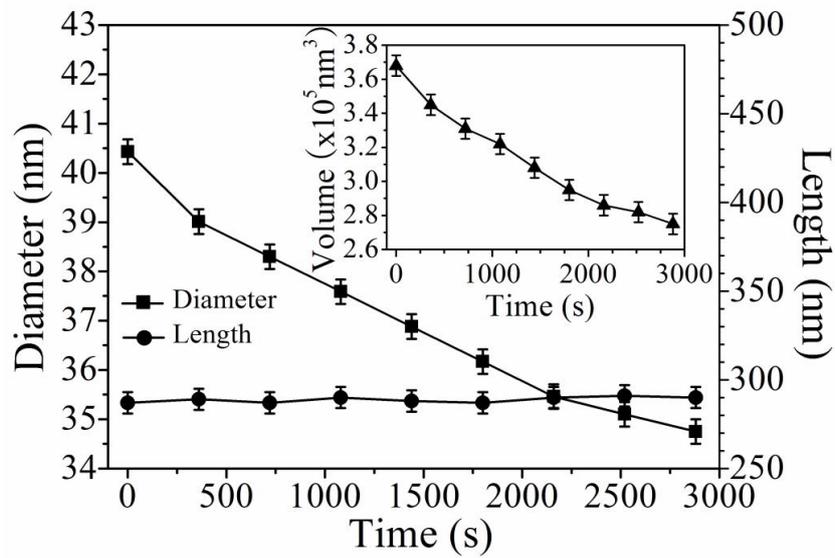

(b)

**Fig. 1** (a) Sequential *in-situ* TEM micrographs showing the typical structure changes and evolution of an axially straight a-SiOx NW as induced by uniform e-beam irradiation with a current density of 1 A/cm²; (b) Evolution of the diameter, length and volume of the axially straight NW segment as shown in (a) against the irradiation time.

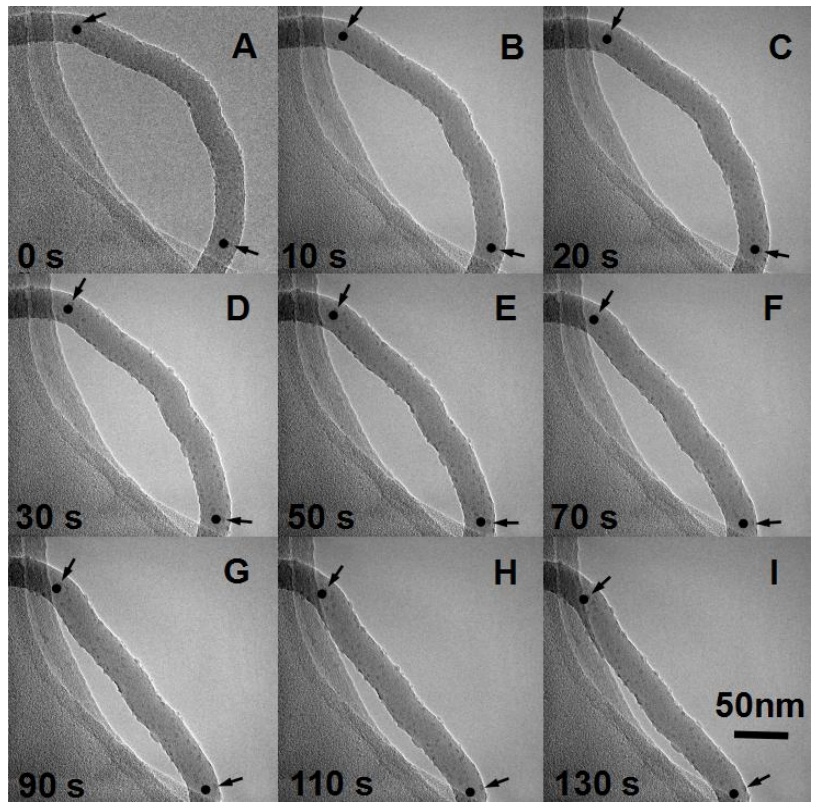

(a)

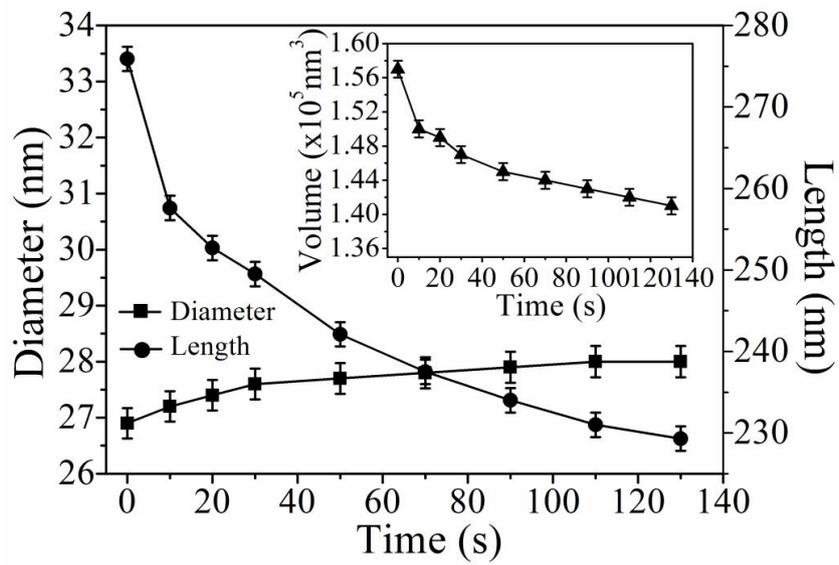

(b)

**Fig. 2** (a) Sequential *in-situ* TEM micrographs showing the typical structure changes and evolution of an axially curved a-SiO$_x$ NW under the same irradiation condition as that in Fig. 1; (b) Evolution of the diameter, length and volume of the axially curved NW segment as shown in (a) against the irradiation time.

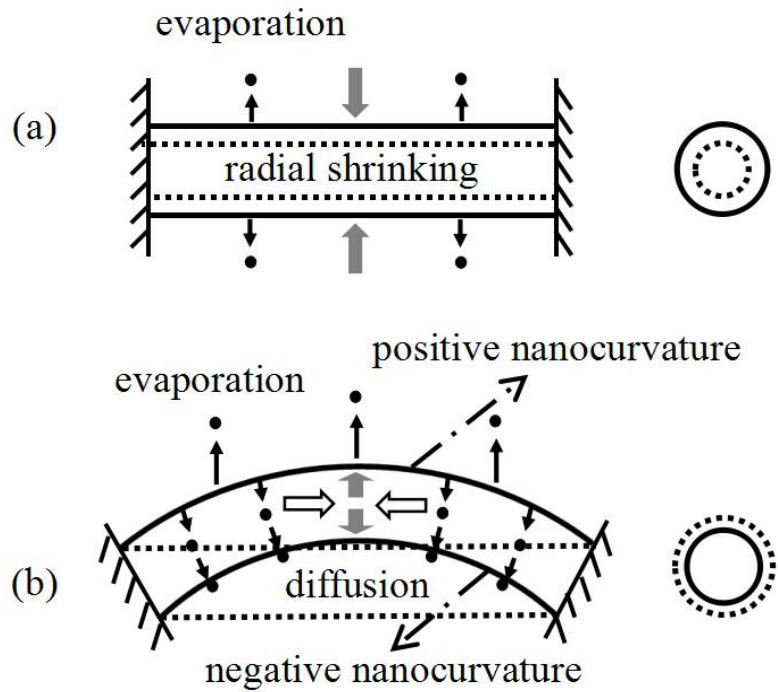

**Fig. 3** Schematic diagrams showing the athermal diffusion or plastic flow and athermal evaporation of wire atoms on axially straight (a) and axially curved (b) NW segments and their resulting structure changes.